\documentclass[12 pt,twoside,aps,prd,amsmath,amssymb,
tightenlines,showpacs,showkeys,eqsecnum]{revtex4}

\usepackage{graphicx}
\usepackage{mathptmx}
\usepackage{bm}
\newcommand {\pr}{\partial}
\newcommand {\cG}{\cal G}
\newcommand {\cD}{\cal D}
\newcommand {\bg}{\bar \gamma}
\newcommand {\G}{\Gamma}
\newcommand {\bp}{\bar \psi}
\newcommand {\p}{\psi}

\def\myfigure #1#2#3#4
{\begin{figure}[ht]\begin{center}
\includegraphics[width=#2 \textwidth]{#1.eps}
\parbox[t]{#4\textwidth}{\caption{#3}\label{#1}}
\end{center}\end{figure}}

\def \myfigures #1#2#3#4#5#6#7#8
{\begin{figure}[ht]
    \begin{center}
        \includegraphics[width=#2 \textwidth]{#1.eps}
        \hfill
        \includegraphics[width=#6 \textwidth]{#5.eps}
        \parbox[t]{#4\textwidth}{\caption {#3}\label{#1}}
        \hfill
        \parbox[t]{#8\textwidth}{\caption {#7}\label{#5}}
    \end{center}
\end{figure} }

\begin{document}
\baselineskip -24pt
\title{Nonlinear Spinor Fields in Bianchi type-VI space-time}
\author{Bijan Saha}
\affiliation{Laboratory of Information Technologies\\
Joint Institute for Nuclear Research\\
141980 Dubna, Moscow region, Russia} \email{bijan@jinr.ru}
\homepage{http://bijansaha.narod.ru}

\begin{abstract}

Within the scope of Bianchi type-VI cosmological model we study the
role of spinor field in the evolution of the Universe. It is found
that due to the spinor affine connections the energy-momentum tensor
of the spinor field possesses non-diagonal components. The
non-triviality of non-diagonal components of the energy-momentum
tensor imposes some severe restrictions either on the spinor field
or on the metric functions or on both of them. But unlike in cases
of Bianchi type-I or $VI_0$, in case of Bianchi type-VI model it
does not lead to the elimination of spinor field nonlinearity or
mass term in the spinor field Lagrangian. It is also found that
depending on the sign of self-coupling constant the model can give
rise to late time acceleration or generate oscillatory mode of
evolution.
\end{abstract}



\maketitle

\bigskip

\section{Introduction}

With the more and more observational data available from far sky,
the need for a change in the standard cosmological paradigm becomes
inevitable. Prior to 1998 when we had no idea about the accelerating
mode of expansion the available observational data were well fit in
the standard Einstein model. But the discovery and further
reconfirmation of the existence of the late time accelerated mode of
expansion \cite{riess,perlmutter} have opened a new window for
change. Along with that comes out a number of alternative models of
the evolution of the Universe.

The most popular among the models are those which consider the old
Einstein theory with a new "matter" as a source field.  The models
with $\Lambda$-term \cite{starobinsky,PRpadma,2006APSS302-83-91},
quintessence
\cite{Carden,chimento1,Linder1,olivares,zlatev,2005ChineseJPhys43-1035-1043},
Chaplygin gas \cite{Kamenshchik,Amendola,Bean,bento,B1,B2,B4,bilic}
etc. are among the most studied ones, though some other models of
dark energy are also proposed. After some remarkable works by
different authors
\cite{henneaux,ochs,saha1997a,saha1997b,saha2001a,saha2004a,
saha2004b,saha2006c,saha2006e,saha2007,saha2006d,greene,ribas,souza,kremer},
showing the important role of spinor field in the evolution of the
Universe, it has been extensively used to model the dark energy.
This success is directly related to its ability to answer some
fundamental questions of modern cosmology: (i) Problem of initial
singularity and its possible elimination
\cite{saha1997a,saha1997b,saha2001a,saha2004a,saha2004b,PopPLB,PopPRD,PopGREG,FabIJTP};
(ii) problem of isotropization
\cite{misner,saha2001a,saha2004a,saha2006c,PopPRD} and (iii) late
time acceleration of the Universe \cite{ribas,
saha2006d,saha2006e,saha2007,PopGREG,PopPLB,FabIJTP,ELKO,FabJMP,PopPRD}.
Moreover recently it was found that the spinor field can also
describe the different characteristics of matter from ekpyrotic
matter to phantom matter, as well as Chaplygin gas
\cite{krechet,saha2010a,saha2010b,saha2011,saha2012}.

It should be noticed that in earlier works only the diagonal
components of the energy-momentum tensor of the spinor field were
taken into account. But recently  it was shown that due to its
specific behavior in curve spacetime the spinor field can
significantly change not only the geometry of spacetime but itself
as well. The existence of nontrivial non-diagonal components of the
energy-momentum tensor plays a vital role in this matter. In
\cite{sahaIJTP2014,sahaAPSS2015} it was shown that depending on the
type restriction imposed on the non-diagonal components of the
energy-momentum tensor, the initially Bianchi type-I evolves into a
LRS Bianchi type-I spacetime or FRW one from the very beginning,
whereas the model may describe a general Bianchi type-I spacetime
but in that case the spinor field becomes massless and linear. The
same thing happens for a Bianchi type-$VI_0$ spacetime, i.e., the
geometry of Bianchi type-$VI_0$ spacetime does not allow the
existence of a massive and/or nonlinear spinor field
\cite{sahabvi0}.

Anisotropic Bianchi type VI cosmological models were studied by many
authors \cite{RJPSV2010,APSSZS2013,IJTPS2013,ECAYA2014,ECAYA2015}.
In this report we study the role of spinor field in the evolution of
a Bianchi type VI anisotropic cosmological model.

\section{Basic equation}

Let us consider the case when the anisotropic space-time is filled
with nonlinear spinor field. The corresponding action can be given
by
\begin{equation}
{\cal S}(g; \psi, \bp) = \int\, L \sqrt{-g} d\Omega \label{action}
\end{equation}
with
\begin{equation}
L= L_{\rm g} + L_{\rm sp}. \label{lag}
\end{equation}
Here $L_{\rm g}$ corresponds to the gravitational field
\begin{equation}
L_{\rm g} = \frac{R}{2\kappa}, \label{lgrav}
\end{equation}
where $R$ is the scalar curvature, $\kappa = 8 \pi G$, with G being
Einstein's gravitational constant and $L_{\rm sp}$ is the spinor
field Lagrangian.

\subsection{Gravitational field}

The gravitational field in our case is given by a Bianchi type-VI
anisotropic space time:

\begin{equation}
ds^2 = dt^2 - a_1^2 e^{-2mx_3} dx_1^2 - a_2^2 e^{2nx_3} dx_2^2 -
a_3^2 dx_3^2, \label{bvi}
\end{equation}
with $a_1,\,a_2$ and $a_3$ being the functions of time only and $m$
and $n$ are some arbitrary constants.

The nontrivial Christoffel symbols for \eqref{bvi} are
\begin{eqnarray}
\G_{01}^{1} &=& \frac{\dot{a_1}}{a_1},\quad \G_{02}^{2} =
\frac{\dot{a_2}}{a_2},\quad
\G_{03}^{3} = \frac{\dot{a_3}}{a_3}, \nonumber\\
\G_{11}^{0} &=& a_1 \dot{a_1} e^{-2mx_3},\quad \G_{22}^{0} = a_2
\dot{a_2} e^{2nx_3},\quad
\G_{33}^{0} = a_3 \dot{a_3},\label{Chrysvi}\\
\G_{31}^{1} &=& -m,\quad \G_{32}^{2} = n,\quad \G_{11}^{3} = \frac{m
a_1^2}{a_3^2} e^{-2mx_3},\quad \G_{22}^{3} = -\frac{n a_2^2}{a_3^2}
e^{2nx_3}. \nonumber
\end{eqnarray}

The nonzero components of the Einstein tensor corresponding to the
metric \eqref{bvi} are
\begin{subequations}
\label{ET}
\begin{eqnarray}
G_1^1 &=&  -\frac{\ddot a_2}{a_2} - \frac{\ddot a_3}{a_3} -
\frac{\dot a_2}{a_2}\frac{\dot a_3}{a_3} + \frac{n^2}{a_3^2}, \label{ET11}\\
G_2^2 &=&  -\frac{\ddot a_3}{a_3} - \frac{\ddot a_1}{a_1} -
\frac{\dot a_3}{a_3}\frac{\dot a_1}{a_1} + \frac{m^2}{a_3^2}, \label{ET22}\\
G_3^3 &=&  -\frac{\ddot a_1}{a_1} - \frac{\ddot a_2}{a_2} -
\frac{\dot a_1}{a_1}\frac{\dot a_2}{a_2} - \frac{m n}{a_3^2}, \label{ET33}\\
G_0^0 &=&  -\frac{\dot a_1}{a_1}\frac{\dot a_2}{a_2} - \frac{\dot
a_2}{a_2}\frac{\dot a_3}{a_3} - \frac{\dot a_3}{a_3}\frac{\dot
a_1}{a_1} + \frac{m^2 - m n + n^2}{a_3^2}, \label{ET00}\\
G_3^0 &=& (m - n) \frac{\dot a_3}{a_3} -  m \frac{\dot a_1}{a_1} + n
 \frac{\dot a_2}{a_2}. \label{ET03}
\end{eqnarray}
\end{subequations}

\subsection{Spinor field}

For a spinor field $\p$, the symmetry between $\p$ and $\bp$ appears
to demand that one should choose the symmetrized Lagrangian
\cite{kibble}. Keeping this in mind we choose the spinor field
Lagrangian as \cite{saha2001a}:

\begin{equation}
L_{\rm sp} = \frac{\imath}{2} \left[\bp \gamma^{\mu} \nabla_{\mu}
\psi- \nabla_{\mu} \bar \psi \gamma^{\mu} \psi \right] - m_{\rm sp}
\bp \psi - F, \label{lspin}
\end{equation}
where the nonlinear term $F$ describes the self-interaction of a
spinor field and can be presented as some arbitrary functions of
invariants generated from the real bilinear forms of a spinor field.
Since $\psi$ and $\psi^{\star}$ (complex conjugate of $\psi$) have
four component each, one can construct $4\times 4 = 16$ independent
bilinear combinations. They are
\begin{subequations}
\label{bf}
\begin{eqnarray}
 S&=& \bar \psi \psi\qquad ({\rm scalar}),   \\
  P&=& \imath \bar \psi \gamma^5 \psi\qquad ({\rm pseudoscalar}), \\
 v^\mu &=& (\bar \psi \gamma^\mu \psi) \qquad ({\rm vector}),\\
 A^\mu &=&(\bar \psi \gamma^5 \gamma^\mu \psi)\qquad ({\rm pseudovector}), \\
Q^{\mu\nu} &=&(\bar \psi \sigma^{\mu\nu} \psi)\qquad ({\rm
antisymmetric\,\,\, tensor}),
\end{eqnarray}
\end{subequations}
where $\sigma^{\mu\nu}\,=\,(\imath/2)[\gamma^\mu\gamma^\nu\,-\,
\gamma^\nu\gamma^\mu]$. Invariants, corresponding to the bilinear
forms, are
\begin{subequations}
\label{invariants}
\begin{eqnarray}
I &=& S^2, \\
J &=& P^2, \\
I_v &=& v_\mu\,v^\mu\,=\,(\bar \psi \gamma^\mu \psi)\,g_{\mu\nu}
(\bar \psi \gamma^\nu \psi),\\
I_A &=& A_\mu\,A^\mu\,=\,(\bar \psi \gamma^5 \gamma^\mu
\psi)\,g_{\mu\nu}
(\bar \psi \gamma^5 \gamma^\nu \psi), \\
I_Q &=& Q_{\mu\nu}\,Q^{\mu\nu}\,=\,(\bar \psi \sigma^{\mu\nu}
\psi)\, g_{\mu\alpha}g_{\nu\beta}(\bar \psi \sigma^{\alpha\beta}
\psi).
\end{eqnarray}
\end{subequations}

According to the Fierz identity,  among the five invariants only $I$
and $J$ are independent as all others can be expressed by them: $I_v
= - I_A = I + J$ and $I_Q = I - J.$ Therefore, we choose the
nonlinear term $F$ to be the function of $I$ and $J$ only, i.e., $F
= F(I, J)$, thus claiming that it describes the nonlinearity in its
most general form. Indeed, without losing generality we can choose
$F = F(K)$, with $K$ taking one of the following expressions
$\{I,\,J,\,I+J,\,I-J\}$.. Here $\nabla_\mu$ is the covariant
derivative of spinor field:
\begin{equation}
\nabla_\mu \psi = \frac{\partial \psi}{\partial x^\mu} -\G_\mu \psi,
\quad \nabla_\mu \bp = \frac{\partial \bp}{\partial x^\mu} + \bp
\G_\mu, \label{covder}
\end{equation}
with $\G_\mu$ being the spinor affine connection. In \eqref{lspin}
$\gamma$'s are the Dirac matrices in curve space-time and obey the
following algebra
\begin{equation}
\gamma^\mu \gamma^\nu + \gamma^\nu \gamma^\mu = 2 g^{\mu\nu}
\label{al}
\end{equation}
and are connected with the flat space-time Dirac matrices $\bg$ in
the following way
\begin{equation}
 g_{\mu \nu} (x)= e_{\mu}^{a}(x) e_{\nu}^{b}(x) \eta_{ab},
\quad \gamma_\mu(x)= e_{\mu}^{a}(x) \bg_a \label{dg}
\end{equation}
where $e_{\mu}^{a}$ is a set of tetrad 4-vectors.

For the metric \eqref{bvi} we choose the tetrad as follows:

\begin{equation}
e_0^{(0)} = 1, \quad e_1^{(1)} = a_1 e^{-mx_3}, \quad e_2^{(2)} =
a_2 e^{nx_3}, \quad e_3^{(3)} = a_3. \label{tetradvi}
\end{equation}

The Dirac matrices $\gamma^\mu(x)$ of Bianchi type-VI space-time are
connected with those of Minkowski one as follows:
$$ \gamma^0=\bg^0,\quad \gamma^1 = \frac{ e^{m x_3}}{a_1} \bg^1,
\quad \gamma^2= \frac{ e^{-n x_3}}{a_2}\bg^2,\quad \gamma^3 = \frac{
1}{a_3}\bg^3$$

$$\gamma^5 = - \imath \sqrt{-g}
\gamma^0\gamma^1\gamma^2\gamma^3 = - \imath \bg^0\bg^1\bg^2\bg^3 =
\bg^5
$$
with
\begin{eqnarray}
\bg^0\,=\,\left(\begin{array}{cc}I&0\\0&-I\end{array}\right), \quad
\bg^i\,=\,\left(\begin{array}{cc}0&\sigma^i\\
-\sigma^i&0\end{array}\right), \quad
\gamma^5 = \bg^5&=&\left(\begin{array}{cc}0&-I\\
-I&0\end{array}\right),\nonumber
\end{eqnarray}
where $\sigma_i$ are the Pauli matrices:
\begin{eqnarray}
\sigma^1\,=\,\left(\begin{array}{cc}0&1\\1&0\end{array}\right),
\quad \sigma^2\,=\,\left(\begin{array}{cc}0&-\imath\\
\imath&0\end{array}\right), \quad
\sigma^3\,=\,\left(\begin{array}{cc}1&0\\0&-1\end{array}\right).
\nonumber
\end{eqnarray}
Note that the $\bg$ and the $\sigma$ matrices obey the following
properties:
\begin{eqnarray}
\bg^i \bg^j + \bg^j \bg^i = 2 \eta^{ij},\quad i,j = 0,1,2,3
\nonumber\\
\bg^i \bg^5 + \bg^5 \bg^i = 0, \quad (\bg^5)^2 = I,
\quad i=0,1,2,3 \nonumber\\
\sigma^j \sigma^k = \delta_{jk} + i \varepsilon_{jkl} \sigma^l,
\quad j,k,l = 1,2,3 \nonumber
\end{eqnarray}
where $\eta_{ij} = \{1,-1,-1,-1\}$ is the diagonal matrix,
$\delta_{jk}$ is the Kronekar symbol and $\varepsilon_{jkl}$ is the
totally antisymmetric matrix with $\varepsilon_{123} = +1$.

The spinor affine connection matrices $\G_\mu (x)$ are uniquely
determined up to an additive multiple of the unit matrix by the
equation
\begin{equation}
\frac{\pr \gamma_\nu}{\pr x^\mu} - \G_{\nu\mu}^{\rho}\gamma_\rho -
\G_\mu \gamma_\nu + \gamma_\nu \G_\mu = 0, \label{afsp}
\end{equation}
with the solution
\begin{equation}
\Gamma_\mu = \frac{1}{4} \bg_{a} \gamma^\nu \partial_\mu e^{(a)}_\nu
- \frac{1}{4} \gamma_\rho \gamma^\nu \Gamma^{\rho}_{\mu\nu}.
\label{sfc}
\end{equation}

From the Bianchi type-VI metric \eqref{sfc} one finds the following
expressions for spinor affine connections:
\begin{subequations}
\label{sac123}
\begin{eqnarray}
\G_0 &=& 0, \label{sac0}\\  \G_1 &=& \frac{1}{2}\left(\dot a_1
\bg^1\bg^0 - m\frac{a_1}{a_3} \bg^1\bg^3\right) e^{-mx_3},
\label{sac1}\\  \G_2 &=& \frac{1}{2}\left(\dot a_2 \bg^2\bg^0 +
n\frac{a_2}{a_3} \bg^2\bg^3\right) e^{nx_3}, \label{sac2}\\  \G_3
&=& \frac{\dot a_3}{2} \bg^3 \bg^0. \label{sac3}
\end{eqnarray}
\end{subequations}

\subsection{Field equations}

Variation of \eqref{action} with respect to the metric function
$g_{\mu \nu}$ gives the Einstein field equation
\begin{equation}
G_\mu^\nu = R_\mu^\nu - \frac{1}{2} \delta_\mu^\nu R = -\kappa
T_\mu^\nu, \label{EEg}
\end{equation}
where $R_\mu^\nu$ and $R$ are the Ricci tensor and Ricci scalar,
respectively. Here $T_\mu^\nu$ is the energy momentum tensor of the
spinor field.

Varying \eqref{lspin} with respect to $\bp (\psi)$ one finds the
spinor field equations:
\begin{subequations}
\label{speq}
\begin{eqnarray}
\imath\gamma^\mu \nabla_\mu \psi - m_{\rm sp} \psi - {\cD} \psi -
 \imath {\cG} \gamma^5 \psi &=&0, \label{speq1} \\
\imath \nabla_\mu \bp \gamma^\mu +  m_{\rm sp} \bp + {\cD}\bp +
\imath {\cG} \bp \gamma^5 &=& 0, \label{speq2}
\end{eqnarray}
\end{subequations}
where we denote ${\cD} = 2 S F_K K_I$ and ${\cG} = 2 P F_K K_J$,
with $F_K = dF/dK$, $K_I = dK/dI$ and $K_J = dK/dJ.$ In view of
\eqref{speq} can be rewritten as
\begin{eqnarray}
L_{\rm sp} & = & \frac{\imath}{2} \left[\bp \gamma^{\mu}
\nabla_{\mu} \psi- \nabla_{\mu} \bar \psi \gamma^{\mu} \psi \right]
- m_{\rm sp} \bp \psi - F(I,\,J)
\nonumber \\
& = & \frac{\imath}{2} \bp [\gamma^{\mu} \nabla_{\mu} \psi - m_{\rm
sp} \psi] - \frac{\imath}{2}[\nabla_{\mu} \bar \psi \gamma^{\mu} +
m_{\rm sp} \bp] \psi
- F(I,\,J),\nonumber \\
& = & 2 (I F_I + J F_J) - F = 2 K F_K - F(K), \label{lspin01}
\end{eqnarray}
where $K = \{I,\,J,\,I + J,\,I - J\}$.

\subsection{Energy momentum tensor of the spinor field}

The energy-momentum tensor of the spinor field is given by
\begin{equation}
T_{\mu}^{\rho}=\frac{\imath}{4} g^{\rho\nu} \left(\bp \gamma_\mu
\nabla_\nu \psi + \bp \gamma_\nu \nabla_\mu \psi - \nabla_\mu \bar
\psi \gamma_\nu \psi - \nabla_\nu \bp \gamma_\mu \psi \right) \,-
\delta_{\mu}^{\rho} L_{\rm sp}. \label{temsp}
\end{equation}

Then in view of \eqref{covder} and \eqref{lspin01} the
energy-momentum tensor of the spinor field can be written as
\begin{eqnarray}
T_{\mu}^{\,\,\,\rho}&=&\frac{\imath}{4} g^{\rho\nu} \left(\bp
\gamma_\mu
\partial_\nu \psi + \bp \gamma_\nu \partial_\mu \psi -
\partial_\mu \bar \psi \gamma_\nu \psi - \partial_\nu \bp \gamma_\mu
\psi \right)\nonumber\\
& - &\frac{\imath}{4} g^{\rho\nu} \bp \left(\gamma_\mu \G_\nu +
\G_\nu \gamma_\mu + \gamma_\nu \G_\mu + \G_\mu \gamma_\nu\right)\psi
 \,- \delta_{\mu}^{\rho} \left(2 K F_K - F(K)\right). \label{temsp0}
\end{eqnarray}
As is seen from \eqref{temsp0}, is case if for a given metric
$\G_\mu$'s are different, there arise nontrivial non-diagonal
components of the energy momentum tensor.

We consider the case when the spinor field depends on $t$ only, i.e.
$\psi = \psi (t)$. Then inserting \eqref{covder} into \eqref{temsp0}
one finds
\begin{subequations}
\label{Ttotext}
\begin{eqnarray}
T_0^0 & = & \frac{\imath}{2} g^{00} \left(\bp \gamma_0
\dot \psi - \dot \bp \gamma_0 \psi\right) - L_{\rm sp},\label{Ttot00ext}\\
T_1^1 & = & -\frac{\imath}{2} g^{11} \bp \left(\gamma_1 \G_1 + \G_1 \gamma_1\right) \psi - L_{\rm sp},\label{Ttot11ext}\\
T_2^2 & = & -\frac{\imath}{2} g^{22} \bp \left(\gamma_2 \G_2 + \G_2 \gamma_2\right) \psi- L_{\rm sp},\label{Ttot22ext}\\
T_3^3 & = & -\frac{\imath}{2} g^{33} \bp \left(\gamma_3 \G_3 + \G_3 \gamma_3\right) \psi- L_{\rm sp},\label{Ttot33ext}\\
T_1^0 & = & \frac{\imath}{4} g^{00} \left(\bp \gamma_1 \dot \psi -
\dot \bp \gamma_1 \psi\right) - \frac{\imath}{4} g^{00} \bp
\left(\gamma_0 \G_1 + \G_1 \gamma_0\right) \psi, \label{Ttot01ext}\\
T_2^0 & = & \frac{\imath}{4} g^{00} \left(\bp \gamma_2 \dot \psi -
\dot \bp \gamma_2 \psi\right) - \frac{\imath}{4} g^{00} \bp
\left(\gamma_0 \G_2 + \G_2 \gamma_0\right) \psi, \label{Ttot02ext}\\
T_3^0 & = & \frac{\imath}{4} g^{00} \left(\bp \gamma_3 \dot \psi -
\dot \bp \gamma_3 \psi\right) - \frac{\imath}{4} g^{00} \bp
\left(\gamma_0 \G_3 + \G_3 \gamma_0\right) \psi, \label{Ttot03ext}\\
T_2^1 & = & -\frac{\imath}{4} g^{11} \bp \left(\gamma_2 \G_1 + \G_1
\gamma_2 + \gamma_1 \G_2 + \G_2 \gamma_1\right) \psi, \label{Ttot12ext}\\
T_3^2 & = & -\frac{\imath}{4} g^{22} \bp \left(\gamma_3 \G_2 + \G_2
\gamma_3 + \gamma_2 \G_3 + \G_3 \gamma_2\right) \psi, \label{Ttot23ext}\\
T_3^1 & = &  -\frac{\imath}{4} g^{11} \bp \left(\gamma_3 \G_1 + \G_1
\gamma_3 + \gamma_1 \G_3 + \G_3 \gamma_1\right) \psi.
\label{Ttot31ext}
\end{eqnarray}
\end{subequations}

Further inserting \eqref{sac123} into \eqref{Ttotext} after a little
manipulations for the components of the energy-momentum tensor one
finds:
\begin{subequations}
\label{Ttot}
\begin{eqnarray}
T_0^0 & = & m_{\rm sp} S + F(K), \label{emt00}\\
T_1^1 &=& T_2^2 = T_3^3 =  F(K) - 2 K F_K, \label{emtii}\\
T_1^0 &=& -\frac{\imath\, n\, e^{-m x_3}}{4}  \frac{a_1}{a_3}\, \bp
\bg^3 \bg^1 \bg^0 \psi
= -\frac{n\, e^{-m x_3}}{4}  \frac{a_1}{a_3} \, A^2 , \label{emt01} \\
T_2^0 &=&-\frac{\imath\, m\, e^{n x_3}}{4}  \frac{a_2}{a_3} \, \bp
\bg^2 \bg^3 \bg^0 \psi
= -\frac{m\, e^{n x_3}}{4} \frac{a_2}{a_3} \,A^1, \label{emt02} \\
T_3^0 &=& 0, \label{emt03} \\
T_2^1 &=& \frac{\imath\, e^{(m +n) x_3}}{4} \frac{a_2}{a_1}
\left[\left(\frac{\dot a_1}{a_1} - \frac{\dot a_2}{a_2}\right) \bp
\bg^1 \bg^2 \bg^0 \psi - \frac{m + n}{a_3} \bp \bg^1 \bg^2 \bg^3
\psi \right] \nonumber \\
&=& \frac{e^{(m +n) x_3}}{4} \frac{a_2}{a_1} \left[\left(\frac{\dot
a_1}{a_1} - \frac{\dot a_2}{a_2}\right) A^3
- \frac{m + n}{a_3}A^0\right] , \label{emt12}\\
T_3^1 &=&\frac{\imath\, e^{m x_3}}{4} \frac{a_3}{a_1}
\left(\frac{\dot a_3}{a_3} - \frac{\dot a_1}{a_1}\right) \bp \bg^3
\bg^1 \bg^0 \psi = \frac{e^{m x_3}}{4} \frac{a_3}{a_1}
\left(\frac{\dot a_3}{a_3} - \frac{\dot a_1}{a_1}\right) A^2 \label{emt13}\\
T_3^2 &=&\frac{\imath\, e^{-n x_3}}{4} \frac{a_3}{a_2}
\left(\frac{\dot a_2}{a_2} - \frac{\dot a_3}{a_3}\right) \bp \bg^2
\bg^3 \bg^0 \psi = \frac{e^{-n x_3}}{4} \frac{a_3}{a_2}
\left(\frac{\dot a_2}{a_2} - \frac{\dot a_3}{a_3}\right)A^1.
\label{emt23}
\end{eqnarray}
\end{subequations}

As one sees from \eqref{Ttotext} and \eqref{Ttot} the non-triviality
of non-diagonal components of the energy momentum tensors is
directly connected with the affine spinor connections $\G_i$'s.

From \eqref{speq} one can write the equations for bilinear spinor
forms \eqref{bf}:
\begin{subequations}
\label{inv}
\begin{eqnarray}
\dot S_0  +  {\cG} A_{0}^{0} &=& 0, \label{S0} \\
\dot P_0  -  \Phi A_{0}^{0} &=& 0, \label{P0}\\
\dot A_{0}^{0} -\frac{m-n}{a_3} A_{0}^{3} +  \Phi P_0 -  {\cG}
S_0 &=& 0, \label{A00}\\
\dot A_{0}^{3} -\frac{m-n}{a_3} A_{0}^{0} &=& 0, \label{A03}\\
\dot v_{0}^{0} - \frac{m-n}{a_3} v_{0}^{3} &=& 0,\label{v00} \\
\dot v_{0}^{3} - \frac{m-n}{a_3} v_{0}^{0} +
\Phi Q_{0}^{30} +  {\cG} Q_{0}^{21} &=& 0,\label{v03}\\
\dot Q_{0}^{30}  -  \Phi v_{0}^{3} &=& 0,\label{Q030} \\
\dot Q_{0}^{21}  -  {\cG} v_{0}^{3} &=& 0, \label{Q021}
\end{eqnarray}
\end{subequations}
where we denote $S_0 = S V,\, P_0 = P V,\, A_0^\mu = A_\mu V,\,
v_0^\mu = v^\mu V,\, Q_0^{\mu \nu} = Q^{\mu \nu} V$ and $\Phi =
m_{\rm sp} + {\cD}$. Here we also introduce the volume scale
\begin{equation}
V = a_1 a_2 a_3. \label{VDef}
\end{equation}

Combining these equations together and taking the first integral one
gets
\begin{subequations}
\label{inv0}
\begin{eqnarray}
(S_{0})^{2} + (P_{0})^{2} + (A_{0}^{0})^{2} - (A_{0}^{3})^{2} &=&
C_1 = {\rm Const}, \label{inv01}\\
(Q_{0}^{30})^{2} + (Q_{0}^{21})^{2} + (v_{0}^{3})^{2} -
(v_{0}^{0})^{2} &=& C_2 = {\rm Const} \label{inv02}
\end{eqnarray}
\end{subequations}

Now let us consider the Einstein field equations. In view of
\eqref{ET} and \eqref{Ttot} we find the following system of Einstein
Equations

\begin{subequations}
\label{EE}
\begin{eqnarray}
\frac{\ddot a_2}{a_2} + \frac{\ddot a_3}{a_3} +
\frac{\dot a_2}{a_2}\frac{\dot a_3}{a_3} - \frac{n^2}{a_3^2} &=& \kappa\left(F(K) - 2 K F_K\right), \label{EE11}\\
\frac{\ddot a_3}{a_3} + \frac{\ddot a_1}{a_1} +
\frac{\dot a_3}{a_3}\frac{\dot a_1}{a_1} - \frac{m^2}{a_3^2} &=& \kappa\left(F(K) - 2 K F_K\right), \label{EE22}\\
\frac{\ddot a_1}{a_1} + \frac{\ddot a_2}{a_2} +
\frac{\dot a_1}{a_1}\frac{\dot a_2}{a_2} + \frac{m n}{a_3^2} &=& \kappa\left(F(K) - 2 K F_K\right), \label{EE33}\\
\frac{\dot a_1}{a_1}\frac{\dot a_2}{a_2} + \frac{\dot
a_2}{a_2}\frac{\dot a_3}{a_3} + \frac{\dot a_3}{a_3}\frac{\dot
a_1}{a_1} - \frac{m^2 - m n + n^2}{a_3^2} &=&   \kappa\left(m_{\rm
sp} S + F(K)\right), \label{EE00}\\
\left(m - n\right) \frac{\dot a_3}{a_3} - m \frac{\dot a_1}{a_1} + n
\frac{\dot a_2}{a_2}  &=& 0, \label{EE03}\\
0 &=& \frac{n\,e^{-m x_3}}{4}  \frac{a_1}{a_3} \, A^2, \label{AC01} \\
0 &=& \frac{m\,e^{n x_3}}{4}  \frac{a_2}{a_3} \,A^1, \label{AC02} \\
0 &=& \frac{e^{(m +n) x_3}}{4} \frac{a_2}{a_1}
\left[\left(\frac{\dot a_1}{a_1} - \frac{\dot a_2}{a_2}\right) A^3
- \frac{m + n}{a_3}A^0\right], \label{AC12}\\
0 &=& \frac{e^{m x_3}}{4} \frac{a_3}{a_1}
\left(\frac{\dot a_3}{a_3} - \frac{\dot a_1}{a_1}\right) A^2, \label{AC13}\\
0 &=&\frac{e^{-n x_3}}{4} \frac{a_3}{a_2}  \left(\frac{\dot
a_2}{a_2} - \frac{\dot a_3}{a_3}\right)A^1. \label{AC23}
\end{eqnarray}
\end{subequations}

From \eqref{AC01} and \eqref{AC02} one dully finds
\begin{equation}
A^2 = 0, \quad {\rm and} \quad A^1 = 0. \label{A12}
\end{equation}
In view of \eqref{A12} the relations \eqref{AC13} and \eqref{AC23}
fulfill even without imposing restrictions on the metric functions.
From \eqref{AC12} one finds the following relations between $A^0$
and $A^3$:
\begin{equation}
\left(\frac{\dot a_1}{a_1} - \frac{\dot a_2}{a_2}\right) A^3 =
\frac{m + n}{a_3}A^0. \label{A0A3}
\end{equation}
Inserting \eqref{A0A3} into \eqref{A03} one finds
\begin{equation}
\frac{m + n}{m - n}\frac{\dot A_{0}^{3}}{A_{0}^{3}} =
\left(\frac{\dot a_1}{a_1} - \frac{\dot a_2}{a_2}\right),
\label{A3a12}
\end{equation}
with the solution
\begin{equation}
\left({A_{0}^{3}}\right)^{\frac{m + n}{m - n}} = X_{03} \left(\frac{
a_1}{a_2}\right), \quad X_{03} = {\rm const.} \label{A03a12}
\end{equation}

On the other hand from \eqref{EE03} one finds the following relation
between the metric functions
\begin{equation}
a_3 = X_0 \left(\frac{a_1^m}{a_2^n}\right)^{1/(m-n)}. \label{abcrel}
\end{equation}

Thus the non-diagonal components of Einstein equations not only
connected the different metric functions as was found in
\cite{saha2004b}, but also imposes some restrictions on the
components of the spinor field.

To find the metric functions explicitly we have to address the
diagonal components of Einstein system. Explicit presence of $a_3$
force us to impose some additional conditions. In an early work
\cite{saha2004b} we propose two different situations, namely, set
$a_3 = \sqrt{V}$ and $a_3 = V$ which allows us to obtain exact
solutions for the metric functions.

In a recent paper we imposed the proportionality condition, widely
used in literature. Demanding that the expansion is proportion to a
component of the shear tensor, namely
\begin{equation}
\vartheta = N_3 \sigma_3^3.\label{propconvi}
\end{equation}
The motivation behind assuming this condition is explained with
reference to  Thorne \cite{thorne67}. The observations of the
velocity-red-shift relation for extragalactic sources suggest that
Hubble expansion of the universe is isotropic today within $\approx
30$ per cent \cite{kans66,ks66}. To put more precisely, red-shift
studies place the limit
\begin{equation}
\frac{\sigma}{H} \leq 0.3, \label{propconviexp}
\end{equation}
on the ratio of shear $\sigma$ to Hubble constant $H$ in the
neighborhood of our Galaxy today. Collins et al. \cite{Collins} have
pointed out that for spatially homogeneous metric, the normal
congruence to the homogeneous expansion satisfies the condition
$\frac{\sigma}{\theta}$ is constant. Under this proportionality
condition it was also found that the energy-momentum distribution of
the model is strictly isotropic, which is absolutely true for our
case.

Let us now find expansion and shear for BVI metric. The expansion is
given by
\begin{equation}
\vartheta = u^\mu_{;\mu} = u^\mu_{\mu} + \G^\mu_{\mu\alpha}
u^\alpha, \label{expansion}
\end{equation}
and the shear is given by
\begin{equation}
\sigma^2 = \frac{1}{2} \sigma_{\mu\nu} \sigma^{\mu\nu},
\label{shear}
\end{equation}
with
\begin{equation}
\sigma_{\mu\nu} = \frac{1}{2}\left[u_{\mu;\alpha} P^\alpha_\nu +
u_{\nu;\alpha} P^\alpha_\mu \right] - \frac{1}{3} \vartheta
P_{\mu\nu}, \label{shearcomp}
\end{equation}
where the projection vector $P$:
\begin{equation}
P^2 = P, \quad P_{\mu\nu} = g_{\mu\nu} - u_\mu u_\nu, \quad
P^\mu_\nu = \delta^\mu_\nu - u^\mu u_\nu. \label{proj}
\end{equation}
In comoving system we have $u^\mu = (1,0,0,0)$. In this case one
finds
\begin{equation}
\vartheta = \frac{\dot a_1}{a_1} + \frac{\dot a_2}{a_2} + \frac{\dot
a_3}{a_3} = \frac{\dot V}{V}, \label{expbvi}
\end{equation}
and
\begin{subequations}
\label{shearcomps}
\begin{eqnarray}
\sigma_{1}^{1} &=& -\frac{1}{3}\left(-2\frac{\dot a_1}{a_1} +
\frac{\dot
a_2}{a_2} + \frac{\dot a_3}{a_3}\right) =  \frac{\dot a_1}{a_1} - \frac{1}{3} \vartheta, \label{sh11}\\
\sigma_{2}^{2} &=& -\frac{1}{3}\left(-2\frac{\dot a_2}{a_2} +
\frac{\dot a_3}{a_3} +
\frac{\dot a_1}{a_1}\right) =  \frac{\dot a_2}{a_2} - \frac{1}{3} \vartheta, \label{sh22}\\
\sigma_{3}^{3} &=& -\frac{1}{3}\left(-2\frac{\dot a_3}{a_3} +
\frac{\dot a_1}{a_1} + \frac{\dot a_2}{a_2}\right) =  \frac{\dot
a_3}{a_3} - \frac{1}{3} \vartheta. \label{sh33}
\end{eqnarray}
\end{subequations}

One then finds
\begin{equation}
\sigma^ 2 = \frac{1}{2}\left[\sum_{i=1}^3 \left(\frac{\dot
a_i}{a_i}\right)^2 - \frac{1}{3}\vartheta^2\right] =
\frac{1}{2}\left[\sum_{i=1}^3 H_i^2 - \frac{1}{3}\vartheta^2\right].
\label{sheargen}
\end{equation}

Inserting \eqref{abcrel} into \eqref{expbvi}, \eqref{shearcomps} and
\eqref{sheargen} we find

\begin{equation}
\vartheta = \frac{2m - n}{m - n} \frac{\dot a_1}{a_1} +  \frac{m -
2n}{m - n}\frac{\dot a_2}{a_2}, \label{expbvi1}
\end{equation}
and
\begin{subequations}
\label{shearcomps0}
\begin{eqnarray}
\sigma_{1}^{1} &=& \frac{m - 2n}{3(m - n)}\left(\frac{\dot a_1}{a_1}
-
\frac{\dot a_2}{a_2}\right), \label{sh110}\\
\sigma_{2}^{2} &=&  \frac{n - 2m }{3(m - n)}\left(\frac{\dot
a_1}{a_1}
- \frac{\dot a_2}{a_2}\right), \label{sh220}\\
\sigma_{3}^{3} &=& \frac{m + n}{3(m - n)}\left(\frac{\dot a_1}{a_1}
- \frac{\dot a_2}{a_2}\right). \label{sh330}
\end{eqnarray}
\end{subequations}

On account of \eqref{abcrel}, \eqref{sh33}, \eqref{VDef} from
\eqref{propconvi} one finds
\begin{subequations}
\label{Metf}
\begin{eqnarray}
a_1 &=& \left[\frac{X_1^{(m-n)/(m-2n)}}{X_0}\,V\right]^{\frac{1}{3} + N_3\frac{m - 2n}{m+n}}, \label{Met1}\\
a_2 &=& X_1 \left[\frac{X_1^{(m-n)/(m-2n)}}{X_0}\,V\right]^{\frac{1}{3}+ N_3\frac{n - 2m}{m+n}},\label{Met2}\\
a_3 &=& X_0 X_1^{-n/(m-n)}
\left[\frac{X_1^{(m-n)/(m-2n)}}{X_0}\,V\right]^{\frac{1}{3} +N_3},
\label{Met3}
\end{eqnarray}
\end{subequations}
where $X_1$ is an integration constant. Further taking into account
that $V = a_1 a_2 a_3$ from \eqref{Metf} one finds
\begin{equation}
X_1^{\frac{m-n}{m-2n} + \frac{m-2n}{m-n}} = 1, \label{X3}
\end{equation}
with either $X_1 = 1$ or $\frac{m-n}{m-2n} + \frac{m-2n}{m-n} = 0$.
Since $(m-n)^2 + (m-2n)^2 \ne 0$, we conclude $X_1 = 1$. Hence for
the metric functions finally we obtain
\begin{eqnarray}
a_1 = \left[\frac{V}{X_0}\right]^{\frac{1}{3} + N_3\frac{m -
2n}{m+n}}, \quad a_2 = \left[\frac{V}{X_0}\right]^{\frac{1}{3}+
N_3\frac{n - 2m}{m+n}},\quad a_3 = X_0
\left[\frac{V}{X_0}\right]^{\frac{1}{3} +N_3}. \label{Metfnew}
\end{eqnarray}

The equation for $V$ can be found from the Einstein Equation
\eqref{ET} which for some manipulation looks
\begin{equation}
\ddot V = 2 (m^2 - mn + n^2) X_0^{2 N_3 - 4/3} V^{1/3 - 2N_3} +
\frac{3 \kappa}{2} \left[m_{\rm sp} S + 2 \left(F(K) - K
F_K\right)\right] V. \label{Vdefein}
\end{equation}
In order to solve \eqref{Vdefein} we have to know the relation
between the spinor and the gravitational fields. Let us first find
those relations for different $K$. Let us recall that  $K$ takes one
of the following expressions $\{I,\,J,\,I+J,\,I-J\}$, with ${\cD} =
2 S F_K K_I$ and ${\cG} = 2 P F_K K_J$.

In case of $K = I$, i.e. ${\cG} = 0$ from \eqref{S0} we duly have
\begin{equation}
\dot S_0 = 0, \label{S0n}
\end{equation}
with the solution
\begin{equation}
K = I = S^2 = \frac{V_0^2}{V^2}, \quad \Rightarrow \quad S =
\frac{V_0}{V}, \quad V_0 = {\rm const.} \label{SV}
\end{equation}
In  this case spinor field can be either massive or massless.

As far as case with $K$ taking one of the expressions
$\{J,\,I+J,\,I-J\}$ that gives $K_J = \pm 1$ is concerned, it can be
solved exactly only for a massless spinor field.

In case of $K = J$, i.e. $ \Phi = {\cD} = 0$ from \eqref{P0} we duly
have
\begin{equation}
\dot P_0 = 0, \label{P0n}
\end{equation}
with the solution
\begin{equation}
K = J = P^2 = \frac{V_0^2}{V^2}, \quad \Rightarrow \quad P =
\frac{V_0}{V}, \quad V_0 = {\rm const.} \label{PV}
\end{equation}

In case of $K = I + J$ the equations \eqref{S0} and \eqref{P0} can
be rewritten as
\begin{subequations}
\begin{eqnarray}
\dot S_0  +  2 P F_K  A_{0}^{0} &=& 0, \label{S0new} \\
\dot P_0  -  2 S F_K  A_{0}^{0} &=& 0, \label{P0new}
\end{eqnarray}
\end{subequations}
which can be rearranged as
\begin{equation}
S_0 \dot S_0 +  P_0 \dot P_0 = \frac{d}{dt}\left( S_0^2 +
P_0^2\right) = \frac{d}{dt}\left(V^2 K\right) = 0, \label{K0}
\end{equation}
with the solution
\begin{equation}
K = \frac{V_0^2}{V^2}, \quad V_0 = {\rm const.} \label{KV}
\end{equation}
Note that one can represent $S$ and $P$ as follows:
\begin{equation}
S =  \frac{V_0}{V} \sin{\theta}, \quad P = \frac{V_0}{V}
\cos{\theta}. \label{KIpJ}
\end{equation}
The term $\theta$ can be determined from \eqref{S0new} or
\eqref{P0new} on account of \eqref{A0A3}, \eqref{A03a12} and
\eqref{Metfnew}.

Finally, for $K = I - J$ the equations \eqref{S0} and \eqref{P0} can
be rewritten as
\begin{subequations}
\begin{eqnarray}
\dot S_0  -  2 P F_K  A_{0}^{0} &=& 0, \label{S0new1} \\
\dot P_0  -  2 S F_K  A_{0}^{0} &=& 0, \label{P0new1}
\end{eqnarray}
\end{subequations}
which can be rearranged as
\begin{equation}
S_0 \dot S_0 -  P_0 \dot P_0 = \frac{d}{dt}\left( S_0^2 -
P_0^2\right) = \frac{d}{dt}\left(V^2 K\right) = 0, \label{K01}
\end{equation}
with the solution
\begin{equation}
K = \frac{V_0^2}{V^2}, \quad V_0 = {\rm const.} \label{KV1}
\end{equation}
As in previous case one can rewrite $S$ and $P$ as follows:
\begin{equation}
S = \frac{V_0}{V} \cosh{\theta}, \quad P = \frac{V_0}{V}
\sinh{\theta}. \label{KImJ}
\end{equation}
Like previous case $\theta$ can be determined from \eqref{S0new1} or
\eqref{P0new1} on account of \eqref{A0A3}, \eqref{A03a12} and
\eqref{Metfnew}.

\section{Solution to the field equations}

In this section we solve the field equations. Let us begin with the
spinor field equations. In view of \eqref{covder} and \eqref{sac123}
the spinor field equation \eqref{speq1} takes the form

\begin{subequations}
\label{SF1}
\begin{eqnarray}
\imath \bg^0 \left(\dot \psi + \frac{1}{2}\frac{\dot V}{V}
\psi\right) - m_{\rm sp} \psi -\frac{m-n}{2 a_3} \bg^3 \psi - {\cD}
\psi -  \imath {\cG} \bg^5 \psi &=&0, \label{speq1p}\\
\imath \left(\dot \bp + \frac{1}{2}\frac{\dot V}{V} \bp\right)\bg^0
+ m_{\rm sp} \bp -\frac{m-n}{2 a_3}\bp \bg^3   + {\cD}  \bp + \imath
{\cG}\bp \bg^5 &=& 0. \label{speq2p}
\end{eqnarray}
\end{subequations}

As we have already mentioned, $\psi$ is a function of $t$ only. We
consider the 4-component spinor field given by
\begin{eqnarray}
\psi = \left(\begin{array}{c} \psi_1\\ \psi_2\\ \psi_3 \\
\psi_4\end{array}\right). \label{psi}
\end{eqnarray}
Denoting $\phi_i =\sqrt{V} \psi_i$ and $\bar X_0 = (m - n) X_0^{N_3
- 2/3}$ from \eqref{SF1} for the spinor field we find we find
\begin{subequations}
\label{speq1pfg}
\begin{eqnarray}
\dot \phi_1 + \imath\, {\Phi} \phi_1 + \left[\imath\, \frac{\bar X_0}{2V^{1/3 + N_3}} + {\cG}\right] \phi_3 &=& 0, \label{ph1}\\
\dot \phi_2 + \imath\, {\Phi} \phi_2 - \left[\imath\, \frac{\bar X_0}{2V^{1/3 + N_3}} - {\cG}\right] \phi_4 &=& 0, \label{ph2}\\
\dot \phi_3 - \imath\, {\Phi} \phi_3 + \left[\imath\, \frac{\bar X_0}{2V^{1/3 + N_3}} -  {\cG}\right] \phi_1 &=& 0, \label{ph3}\\
\dot \phi_4 - \imath\, {\Phi} \phi_4 - \left[\imath\, \frac{\bar
X_0}{2V^{1/3 + N_3}} + {\cG} \right] \phi_2&=& 0. \label{ph4}
\end{eqnarray}
\end{subequations}
Further denoting ${\cal Y} = \frac{\bar X_0}{2V^{1/3 + N_3}}$ we can
write the foregoing system of equation in the form:
\begin{equation}
\dot \phi = A \phi, \label{phi}
\end{equation}
with $\phi = {\rm
col}\left(\phi_1,\,\phi_2,\,\phi_3,\,\phi_4\right)$ and
\begin{equation}
A = \left(\begin{array}{cccc}-\imath\, \Phi &0 & -\imath\, {\cal Y}
- {\cG}& 0 \\ 0&-\imath\, \Phi& 0 &\imath\, {\cal Y} -
{\cG}\\-\imath\, {\cal Y} + {\cG}&0&\imath\, \Phi&0\\0&\imath\,
{\cal Y} + {\cG} & 0 &\imath\, \Phi \end{array}\right). \label{AMat}
\end{equation}
It can be easily found that
\begin{equation}
{\rm det} A = \left(\Phi^2 + {\cal Y}^2 +{\cG}^2\right)^2.
\label{detA}
\end{equation}

The solution to the equation \eqref{phi} can be written in the form
\begin{equation}
\phi(t) = {\rm T exp}\left(-\int_t^{t_1}  A_1 (\tau) d \tau\right)
\phi (t_1), \label{phi1}
\end{equation}
where
\begin{equation}
A = \left(\begin{array}{cccc}-\imath\, {\cD} &0 & -\imath\, {\cal Y}
- {\cG}& 0 \\ 0&-\imath\, {\cD}& 0 &\imath\, {\cal Y} -
{\cG}\\-\imath\, {\cal Y} + {\cG}&0&\imath\, {\cD}&0\\0&\imath\,
{\cal Y} + {\cG} & 0 &\imath\, {\cD} \end{array}\right).
\label{AMat1}
\end{equation}
and $\phi (t_1)$ is the solution at $t = t_1$. As we have already
shown, $K = V_0^2/V^2$ for $K = \{J,\,I+J,\,I-J\}$ with trivial
spinor-mass and $K = V_0^2/V^2$ for $K=I$ for any spinor-mass. Since
our Universe is expanding, the quantities ${\cD}$, ${\cal Y}$ and
${\cG}$ become trivial at large $t$. Hence in case of  $K = I$ with
non-trivial spinor-mass one can assume $\phi (t_1) = {\rm
col}\left(e^{-\imath m_{\rm sp} t_1},\,e^{-\imath m_{\rm sp}
t_1},\,e^{\imath m_{\rm sp} t_1},\,e^{\imath m_{\rm sp}
t_1}\right)$, whereas for other cases with trivial spinor-mass we
have $\phi (t_1) = {\rm
col}\left(\phi_{1}^{0},\,\phi_{2}^{0},\,\phi_{3}^{0},\,\phi_{4}^{0}\right)$
with $\phi_i^0$ being some constants. Here we have used the fact
that $\Phi = m_{\rm sp} + {\cD}.$ The other way to solve the system
\eqref{speq1pfg} is given in \cite{saha2004b}.

As far as equation for $V$, i.e.,  \eqref{Vdefein} is concerned, we
solve it setting $K =  I$ as in this case we can use the mass term
as well.

Assuming
\begin{equation}
F = \sum_{k} \lambda_k I^{n_k} =  \sum_{k} \lambda_k S^{2 n_k}
\label{nonlinearity}
\end{equation}
on account of $S = V_0/V$ we find

\begin{eqnarray}
\ddot V =  \Phi(V), \quad \Phi(V) = {\bar X} V^{1/3 - 2N_3} +
\frac{3 \kappa}{2} \left[m_{\rm sp}\,V_0 + 2 \sum_{k} \lambda_k( 1 -
n_k) V_0^{2n_k} V^{1 - 2n_k}\right], \label{Vdefein1}
\end{eqnarray}
where ${\bar X} =  2\left(m^2 - mn + n^2 \right) X_0^{(2N_3 -
4/3)}$.

Let us now show the existence and uniqueness of the solution of the
Eq. \eqref{Vdefein1}. For this we study the right hand side of the
Eq. \eqref{Vdefein1}, namely we check whether $\Phi(t,\,V)$
satisfies the Lipschitz condition. In doing so let us rewrite
\eqref{Vdefein1} as
\begin{subequations}
\label{VWsys}
\begin{eqnarray}
\dot W &=& \Phi(V), \label{VWsys1}\\
\dot V &=& W. \label{VWsys2}
\end{eqnarray}
\end{subequations}

Let $(t_*,\,V_*)$ and $(t_*,\,W_*)$ be a particular pair of values
assigned to the real variables $(t,\,V)$ and $(t_*,\,W_*)$ such that
within a rectangular domain ${\bf D}$ surrounding the point
$(t_*,\,V_*)$ and defined by the inequalities
\begin{equation}
|t - t_*| \le A_*, \qquad |V - V_*| \le B_*, \label{rect}
\end{equation}
$\Phi(t,\,V)$ is a one valued continuous function of $t$ and $V$.
Indeed, for the other parameter fixed $\Phi(t,\,V)$ is a one valued
continuous function. Recalling that $V = 0$ corresponds to a
space-time singularity and $V$ is essentially non-negative, for any
nontrivial value of $V$ we conclude that $|\Phi(t,\,V)|$ has an
upper bound $M$ in ${\bf D}$. We also define $h = {\rm min}(A_*,
B_*/M)$. If $h < A_*$, upon $t$ we impose the additional condition
$|t - t_*| < h$. Let $(t_1,\,V_1)$ and $(t_2,\,V_2)$ are two points
within ${\bf D}$.

The Lipschitz condition in this case
\begin{eqnarray}
\sqrt{\left[\Phi (t_1,\,V_1) - \Phi (t_2,\,V_2) \right]^2 + (W_1 -
W_2)^2} \le L \sqrt{(V_1 - V_2)^2 + (W_1 - W_2)^2}, \label{lcsys}
\end{eqnarray}
for any $L > 1$ follows from
\begin{eqnarray}
\sqrt{\left[\Phi (t_1,\,V_1) - \Phi (t_2,\,V_2) \right]^2 } \le L
\sqrt{(V_1 - V_2)^2}. \label{lcsys1}
\end{eqnarray}
Here $L$ is a constant. Inserting $\Phi(t,\,V)$ from
\eqref{Vdefein1} into the left hand side of \eqref{lcsys1} we find
\begin{eqnarray}
\left|\Phi (t_1,\,V_1) - \Phi (t_2,\,V_2) \right| &=& \left| {\bar
X} \left(V_1^{1/3 - 2N_3} - V_2^{1/3 - 2N_3}\right) + 3 \kappa
\sum_{k} \lambda_k( 1 - n_k) V_0^{2n_k}  \left[V_1^{1 - 2n_k} -
V_2^{1 - 2n_k}\right] \right| \nonumber\\
&=& \left|(V_2 - V_1)\left[{\bar X_1} {(V^*)}^{-(2/3 + 2 N_3)} +
\sum_{k} \lambda_k^* {(V^*)^{-2n_k}}\right]\right| \label{lcsys2}
\end{eqnarray}
where $V^* \in [V_1,\, V_2]$, \quad $V_1 > 0, \quad V_2 > 0$. Here
we also denote ${\bar X_1} = (1/3 - 2 N_3) {\bar X}$, and
$\lambda_k^* = 2 \kappa (1 - n_k) (1 - 2 n_k) V_0^{2 n_k}
\lambda_k$. Here we used the mean value theorem.

Since $\left[{\bar X_1} {(V)}^{-(2/3 + 2 N_3)} + \sum_{k}
\lambda_k^* {(V)^{-2n_k}}\right]$ is continuous, it possesses a
maximum $L$ in the interval $[V_1,\,V_2]$, such that
\begin{eqnarray}
\left|\Phi (t_1,\,V_1) - \Phi (t_2,\,V_2) \right| \le L \left|V_2 -
V_1\right| \label{lcsys3}
\end{eqnarray}
in the domain ${\bf D}$. Further extending this study to other
domains it can be shown that the condition \eqref{lcsys3} holds in
${\bf D} = \bigcup_{V_1 > 0, V_2 < \infty} [V_1,\,V_2].$

Thus we see that $\Phi(t,\,V)$ is continuous and satisfies Lipschitz
condition in the domain $D$. Hence Eq. \eqref{Vdefein1} admits a
unique continuous solution.

Once the existence and uniqueness of the solution of
\eqref{Vdefein1} is proved, we can carry our study further. The
first integral of \eqref{Vdefein1} is

\begin{eqnarray}
\dot V = \Phi_1(V), \quad \Phi_1(V) = \sqrt{{\bar X_1} V^{(4/3 -
2N_3)} + 3 \kappa \left[m_{\rm sp} V_0 V + \sum_{k} \lambda_k
V_0^{2n_k} V^{2(1 - n_k)} + {\bar C}\right]}, \label{1stint}
\end{eqnarray}
where we denote $ {\bar X_1 } = 6 {\bar X}/(4 - 6N_3)$ and ${\bar
C}$ is the constant of integration. The solution for $V$ can be
written in quadrature as

\begin{eqnarray}
\int \frac{dV}{\sqrt{{\bar X_1} V^{(4/3 - 2N_3)} + 3 \kappa
\left[m_{\rm sp} V_0 V + \sum_{k} \lambda_k V_0^{2n_k} V^{2(1 -
n_k)}\right] + {\bar C} }} =  t + t_0, \label{quadrature}
\end{eqnarray}
with ${\bar C} $ and $t_0$ being some arbitrary constants.

In what follows we solve the Eqn. \eqref{Vdefein1} numerically. In
doing so we determine $\dot V (0)$ from \eqref{1stint} for the given
value of $ V (0)$.

To determine the character of the evolution, let us first study the
asymptotic behavior of the equation \eqref{Vdefein1}. It should be
recalled that we have $K = V_0^2 / V^2$. Since all the physical
quantities constructed from the spinor fields as well as the
invariants of gravitational fields are inverse function of $V$ of
some degree, it can be concluded that at any spacetime point where
the volume scale becomes zero, it is a singular point
\cite{saha2001a}. So we assume at the beginning $V$ was small but
non-zero. Then from \eqref{Vdefein1} we see that at $t \to 0$ the
nonlinear term prevails if  $ n_k = n_1: n_1 > {\rm max}\left[1/2,\,
N_3 + 1/3\right]$. Recalling that we are considering an expanding
Universe, at $t \to \infty$ the volume scale should be quite large.
In that case the nonlinear term prevails over the first term if $n_k
= n_2: n_2 < {\rm min}\left[1/2,\, N_3 + 1/3\right]$. For $n_k =
n_0: 1 - 2n_0 = 0$, i.e. $n_0 = 1/2$ the spinor field nonlinearity
vanishes and the corresponding term becomes equivalent to the
(effective) mass term.

To define whether the model allows decelerated or accelerated mode
of expansion we also plot deceleration parameter $q$ defines as

\begin{equation}
q = - \frac{V \ddot V}{{\dot V}^2}, \label{decel}
\end{equation}
which in view of \eqref{Vdefein1} and \eqref{1stint} can be
rewritten as

\begin{equation}
q = -\frac{V \Phi (V)}{\Phi_1^2 (V)} = - \frac{{\bar X} V^{4/3 -
2N_3} + \frac{3 \kappa}{2} \left[m_{\rm sp} V_0 V +  2 \sum_{k}
\lambda_k( 1 - n_k) V_0^{2n_k} V^{1 - 2n_k}\right]}{{\bar X_1}
V^{(4/3 - 2N_3)} + 3 \kappa \left[m_{\rm sp} V_0 V + \sum_{k}
\lambda_k V_0^{2n_k} V^{2(1 - n_k)}\right] + {\bar C}},
\label{decelnew}
\end{equation}

Now let us see what happens to deceleration parameter as $t \to
\infty$. As we have already established, for $n_2 < 1/2$ and $n_2 <
1/3 + N_3$ the nonlinear tern prevails and in this case we find
\begin{equation}
q \approx - (1 - n_2) < 0, \label{decelnonlin}
\end{equation}
whereas for $N_3 < 1/6$ and $n_2 > 1/3 + N_3$ we have
\begin{equation}
q \approx - \frac{\bar X}{\bar X_1} = - (2/3 - N_3) < 0.
\label{decelfterm}
\end{equation}
Thus we see that in both cases the Universe expands with
acceleration.

It should also be emphasized that for  $n_1 > 1/2$ and  $N_3 > 1/6$
the mass term prevails asymptotically at $t \to \infty$ and the
Universe expands as a quadratic function of time, i.e., $V |_{t \to
\infty} \propto t^2$.

The above analysis shows that the absence of mass term leads to
constant deceleration parameter, while for a time depending
deceleration parameter the presence of a non-zero mass term is
essential.

Let us also see what happens to EoS (equation of state) parameter in
this case. Inserting \eqref{nonlinearity} into \eqref{emt00} and
\eqref{emtii} one finds the  expressions for energy density
$\varepsilon = T_0^0$ and pressure $p = - T_1^1$ (in this particular
case as $T_1^1 = T_2^2 = T_3^3$):

\begin{equation}
\varepsilon = m_{\rm sp} \frac{V_0}{V} + \sum_{k} \lambda_k
\frac{V_0^{2n_k}}{V^{2n_k}}, \quad p = \sum_{k} \lambda_k (2n_k - 1)
\frac{V_0^{2n_k}}{V^{2n_k}}. \label{edpr}
\end{equation}

In view of \eqref{edpr} for the EoS parameter we find

\begin{equation}
W = \frac{p}{\varepsilon} = \frac{\sum_{k} \lambda_k (2n_k - 1)
\frac{V_0^{2n_k}}{V^{2n_k}}}{m_{\rm sp} \frac{V_0}{V} + \sum_{k}
\lambda_k \frac{V_0^{2n_k}}{V^{2n_k}}}, \label{edpr1}
\end{equation}
It can be shown that at the early stage of evolution the EoS
parameter is dominated by the usual matter, while at later stage the
dark energy becomes dominant. Moreover, at the absence of the mass
term the EoS parameter becomes a constant as in that case $W = 2n_k
- 1$, whereas for a non-trivial mass term the EoS parameter is a
variable function of time. Here we have exploited the fact that, at
any concrete stage of evolution, one of the terms of the sum becomes
predominant; hence others can be overlooked.

There might be some question regarding the choice of nonlinearity in
the form \eqref{nonlinearity}. The reason lies on the fact that the
spinor description of different kinds of fluid and dark energy such
as ekpyrotic matter, dust, radiation, quintessence, Chaplygin gas,
phantom matter etc. is in one form or the other is given by the
power law of the invariants of spinor field. While the spinor
description of fluid or dark energy leads to the elimination of mass
term, the choice \eqref{nonlinearity} still allows us to study the
role of spinor mass on the evolution of the Universe. To show this
let us recall that only in case of $K = I = S^2$ we could express
$K$ in terms of $V$ with a non-trivial mass term in the Lagrangian.
So setting $F = F(S)$ from
\begin{equation}
W = \frac{p}{\varepsilon}, \label{barotropic}
\end{equation}
in view of \eqref{emt00} and \eqref{emtii} one finds
\cite{saha2010a,saha2010b,saha2011,saha2012}
\begin{equation}
F = \lambda S^{1+W} - m S = \lambda \frac{V_0^{1+W}}{V^{1+W}} -
m\frac{V_0}{V} \label{nonlinbarotrop}
\end{equation}
that corresponds to dust ($W = 0$), \quad radiation ($W = 1/3)$),
\quad hard Universe ($W \in (1/3,\,1)$), \quad stiff matter ($W =
1$), \quad quintessence ($W \in (-1/3,\,-1)$),\quad cosmological
constant ($W = -1$), \quad phantom matter ($W < -1$), and  ekpyrotic
matter ($ W > 1$), respectively. Inserting \eqref{nonlinbarotrop}
into \eqref{lspin} one finds that the mass term in this case
vanishes, while the spinor field nonlinearity given by
\eqref{nonlinearity} does not. In this case for energy density and
pressure we find $\varepsilon = \lambda V_0^{1+W}/V^{1+W}$ and $p =
\lambda W V_0^{1+W}/V^{1+W}$, respectively. EoS parameter in this
case is a constant by definition, while in absence of the mass term
the deceleration parameter also becomes a constant. Nevertheless one
can use \eqref{nonlinearity} with the trivial mass term in the
Lagrangian and the sum ($\sum_{k}$) in \eqref{nonlinearity} can be
viewed as multi-component source field with $k$ standing for
different types of matter and dark energy such as ekpyrotic matter,
dust, radiation, quintessence, Chaplygin gas, phantom matter etc.

Comparing \eqref{nonlinearity} with \eqref{nonlinbarotrop} one finds
$2n_k = W + 1$. Further setting the value of $W$ for different fluid
and dark energy we find the corresponding value of $n_k$: dust ($n_k
= 1/2$), \quad radiation ($n_k = 2/3)$), \quad hard Universe ($n_k
\in (2/3,\,1)$), \quad stiff matter ($n_k = 1$), \quad quintessence
($n_k \in (0,\,1/3)$),\quad cosmological constant ($n_k = 0$), \quad
phantom matter ($n_k < 0$), and  ekpyrotic matter ($ n_k > 1$),
respectively. It was shown earlier, when $n_k = 1/2$ the
corresponding term can be added to the mass term. So we can conclude
that the term with $n_k = 1/2$ which also describes dust behaves
like a mass term.

One of the principal advantage of using spinor description of source
field lies on the fact that in this case one needs not think about
whether two or more components considered can be separated. To show
that let us write the Biacnhi identity that leads to

\begin{equation}
T_{\mu;\nu}^{\nu} = T_{\mu,\nu}^{\nu} + \G_{\rho \nu}^{\nu}
T_\mu^\rho -   \G_{\mu \nu}^{\rho} T_\rho^\nu = 0, \label{BIdentity}
\end{equation}
which for the metric \eqref{bvi} on account of the components of the
energy-momentum tensor takes the from
\begin{equation}
\dot \varepsilon + \frac{\dot V}{V}\left(\varepsilon + p\right) = 0.
\label{Conserv}
\end{equation}
Inserting $\varepsilon$ and $p$ from \eqref{emt00} and \eqref{emtii}
from \eqref{Conserv} one finds

\begin{equation}
\frac{m_{\rm sp}}{V}\,\, \frac{d}{dt}\left( S V \right) +
\frac{F_K}{ V^2} \,\, \frac{d}{dt}\left( K V^2 \right) = 0.
\label{Conserv1}
\end{equation}

In case of $K = I = S^2$ \eqref{Conserv1} fulfills identically
thanks to \eqref{SV}, i.e., $S V = const.$ and $K V^2 = const.$,
whereas in the case when $K$ takes one of the following expressions
$\{J,\,I+J,\,I-J\}$, for a massless spinor field \eqref{Conserv1}
fulfills identically thanks to \eqref{PV}, \eqref{KV} and
\eqref{KV1}, i.e., $K V^2 = const.$ Hence if we use spinor
description of different fluid and dark energy simulated from
corresponding equation of sate, the Bianchi identity will be
fulfilled identically without invoking any additional condition.

To this end let us solve the equation for $V$ numerically. For
simplicity let us consider system with two components only. In this
case we have

\begin{eqnarray}
\ddot V &=&  \Phi(V), \label{Vdefeinnum1}\\
\Phi(V) &=& {\bar X} V^{1/3 - 2N_3} + \frac{3 \kappa}{2}
\left[m_{\rm sp}\,V_0 + 2 \lambda_1( 1 - n_1) V_0^{2n_1} V^{1 -
2n_1} + 2 \lambda_2( 1 - n_2) V_0^{2n_2} V^{1 - 2n_2}\right],
\nonumber
\end{eqnarray}
with the first integral

\begin{eqnarray}
\dot V &=& \Phi_1(V), \label{1stintnum}\\
\Phi_1^2(V) &=& {\bar X_1} V^{(4/3 - 2N_3)} + 3 \kappa \left[m_{\rm
sp} V_0 V + \lambda_1 V_0^{2n_1} V^{2(1 - n_1)} + \lambda_2
V_0^{2n_2} V^{2(1 - n_2)}  + {\bar C}\right], \nonumber
\end{eqnarray}

Since we are interested in qualitative picture here, so we set the
value of problem parameters very simple. In doing so we set the
value of $N_3,\, n_1$ and $n_2$ in such as way that none of the four
terms in the right hand side of \eqref{Vdefeinnum1} merge with the
others. Beside this we will consider the coupling constants
$\lambda_1$ and $\lambda_2$ with different signs. The initial value
$V(0)$ is taken to be small but non-zero in such a way that the
right hand side of \eqref{1stintnum} remains non-negative. For the
given initial value $\dot V(0)$ is defined from \eqref{1stintnum}.

From \eqref{1stintnum} it can be easily established that only in
case when both $\lambda_1$ and $\lambda_2$ are positive the model
allows ever expanding solution, whereas, if one of the $\lambda_i$'s
is negative, the non-negativity of the expressions under the
square-root imposes some restrictions on the value of $V$. A
negative $\lambda_1$ generates the minimums while the negative
$\lambda_2$ generates the maximums. In case if the minimum occurs at
a negative value of $V$ we have a Universe that expands to some
maximum value and then contracts before ending in a Big Crunch. If
both maximums and minimums are non-zero, we have periodic solution
with no beginning and no end. In both cases $\lambda_2 < 0$. If
$\lambda_2 > 0$ independent to whether $\lambda_1$ is positive or
negative, we have ever expanding solution.

For simplicity we set $m = 1,\, n = 2,\, X_0 = 1,\, V_0 = 1,\,
m_{\rm sp} = 1,\, C_0 = 1,\, \kappa = 1.$ Fixing $N_3 = -1/3$ from
$n_1 > {\rm max} \left[1/2,\, N_3 + 1/3\right]$ we set $n_1 = 3/2$
(ekpyrotic matter) and $n_1 = 2/3$ (radiation), while from $n_2 <
{\rm min}\left[1/2,\, N_3 + 1/3\right]$ we set $n_2 = 1/4$
(quintessence) and $n_2 = -1$ (phantom matter). As far as coupling
constants are concerned we consider two cases with $\lambda_1 = \{1,
-0.001\}$ and $\lambda_2 = \{1, -1\}$. The initial value of $V (0)$
is taken to be $V(0) = 0.01$.

\begin{figure}[ht]
\centering
\includegraphics[height=70mm]{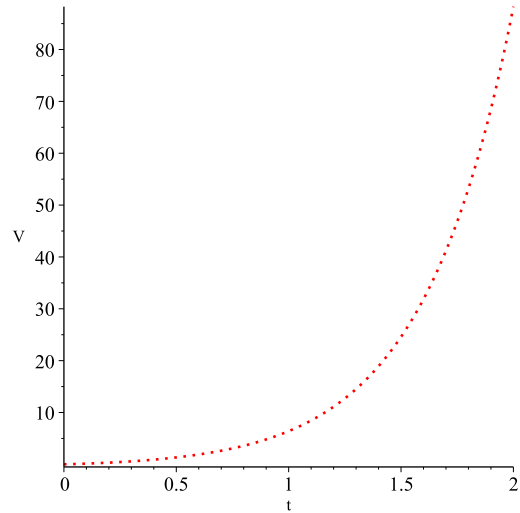} \\
\vskip 1 cm \caption{Evolution of the Universe filled with massive
spinor field  with $n_1 = 2/3$, $n_2 = 1/4$, $\lambda_1 = 1$ and
$\lambda_2 = 1$} \label{BVIVrqpp}.
\end{figure}

\begin{figure}[ht]
\centering
\includegraphics[height=70mm]{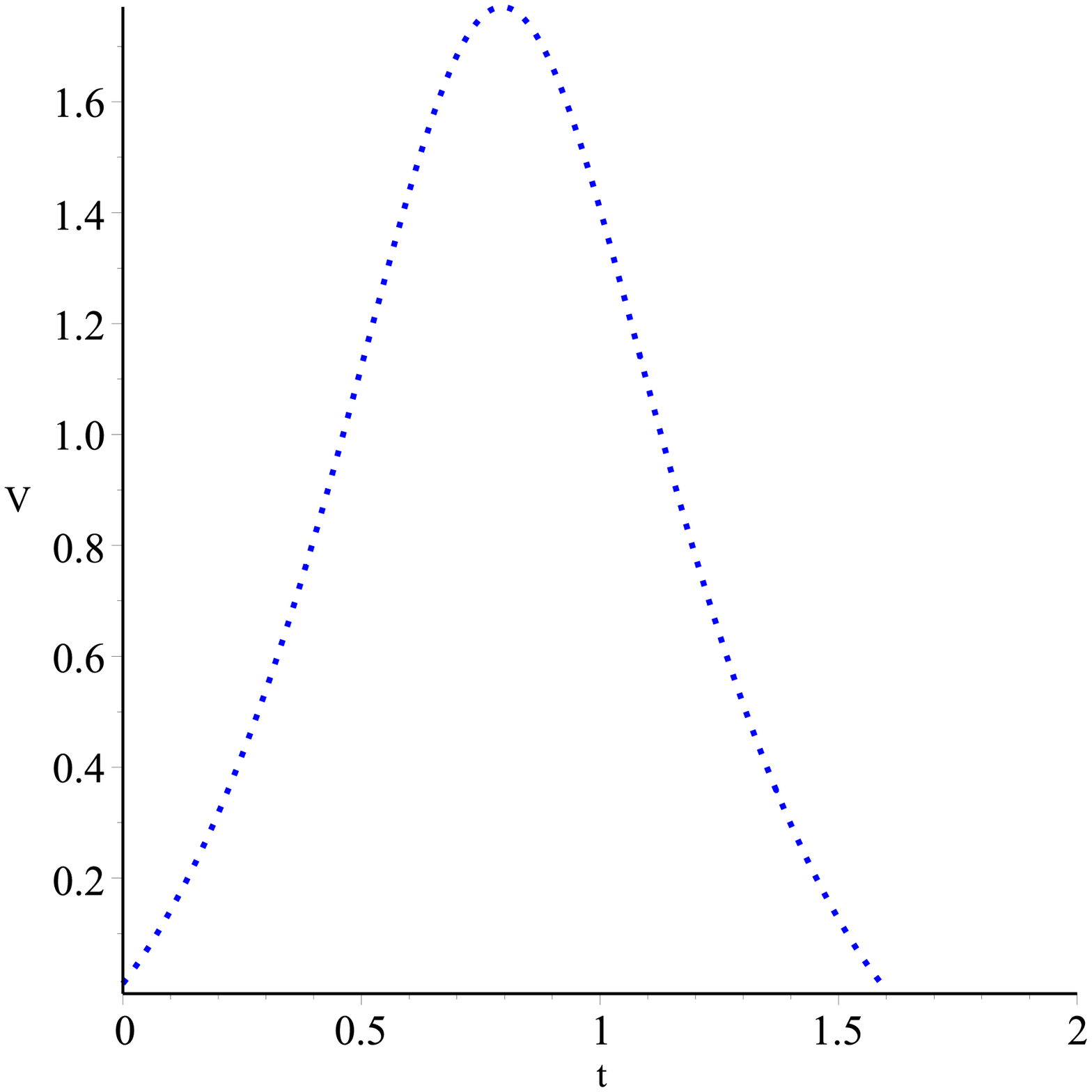} \\ \vskip 1 cm
\caption{Evolution of the Universe filled with massive spinor field
with $n_1 = 2/3$, $n_2 = -1$, $\lambda_1 = 1$ and $\lambda_2 = -1$.}
\label{BVIVrpmpm}.
\end{figure}

\begin{figure}[ht]
\centering
\includegraphics[height=70mm]{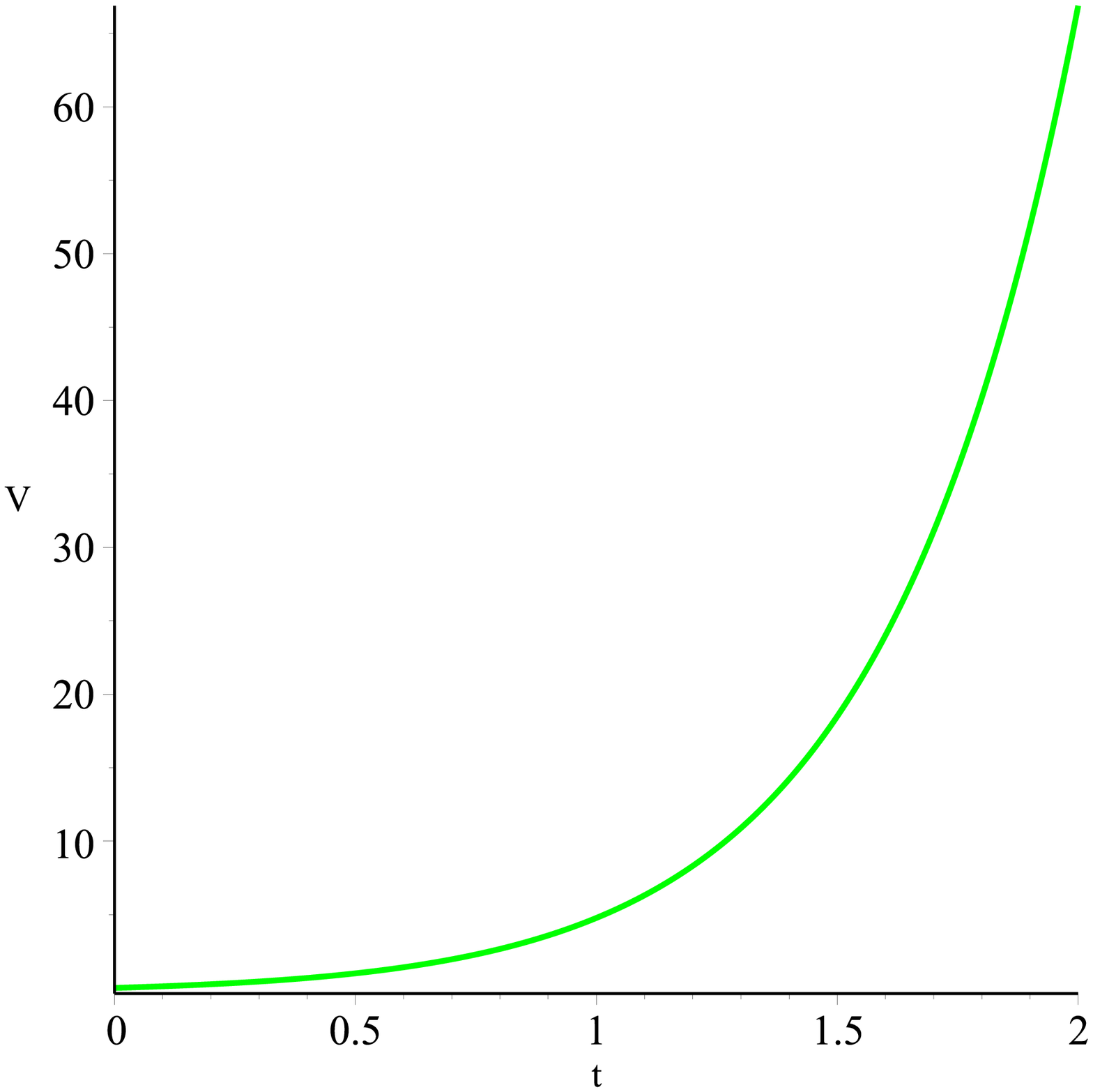} \\
\vskip 1 cm \caption{Evolution of the Universe filled with massive
spinor field  with $n_1 = 3/2$, $n_2 = 1/4$, $\lambda_1 =-0.001$ and
$\lambda_2 = 1$} \label{BVIVemqmp}.
\end{figure}

\begin{figure}[ht]
\centering
\includegraphics[height=70mm]{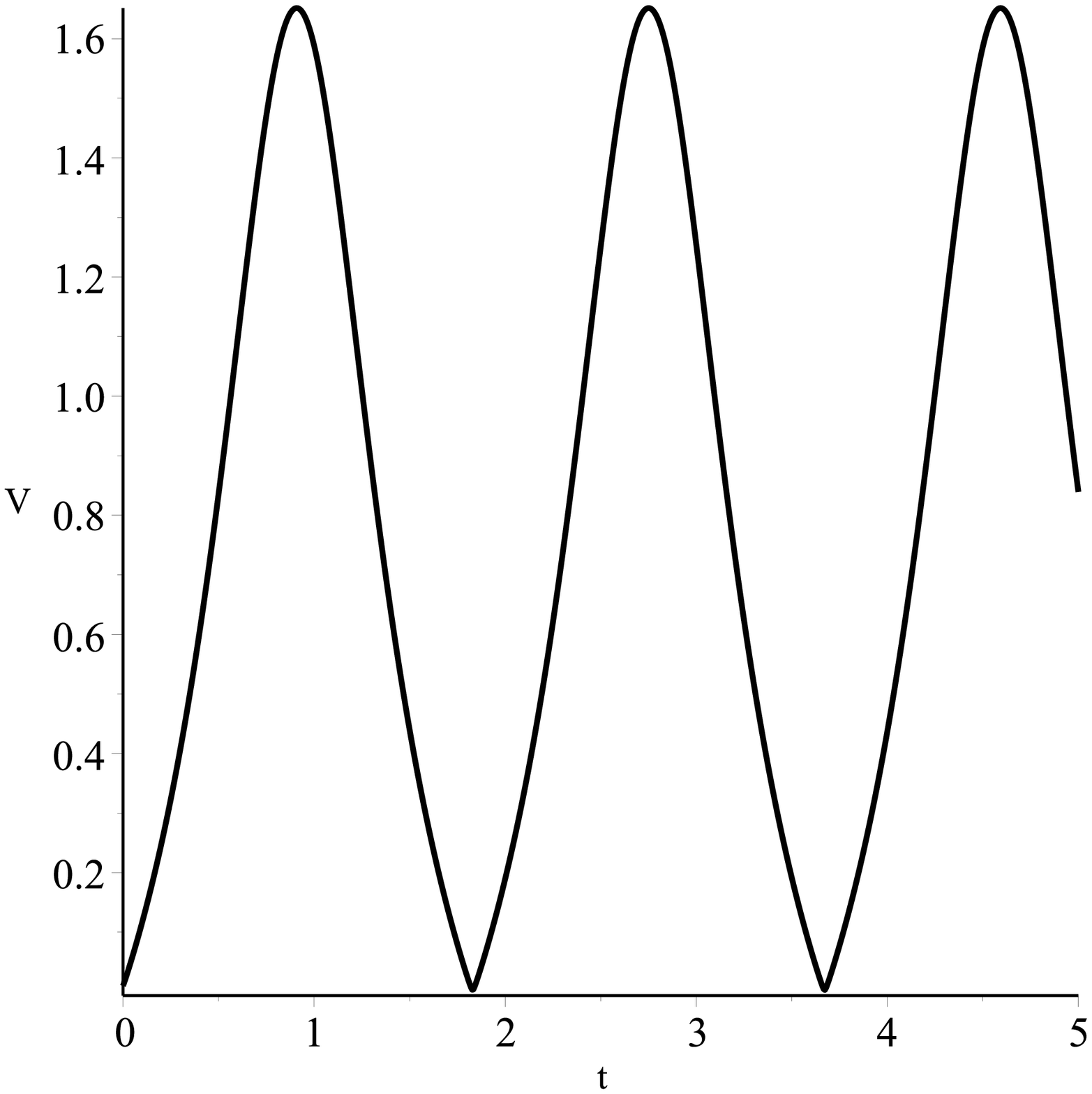} \\ \vskip 1 cm
\caption{Evolution of the Universe filled with massive spinor field
with $n_1 = 2/3$, $n_2 = -1$, $\lambda_1 = -0.001$ and $\lambda_2 =
-1$.} \label{BVIVempmmm}.
\end{figure}

\begin{figure}[ht]
\centering
\includegraphics[height=70mm]{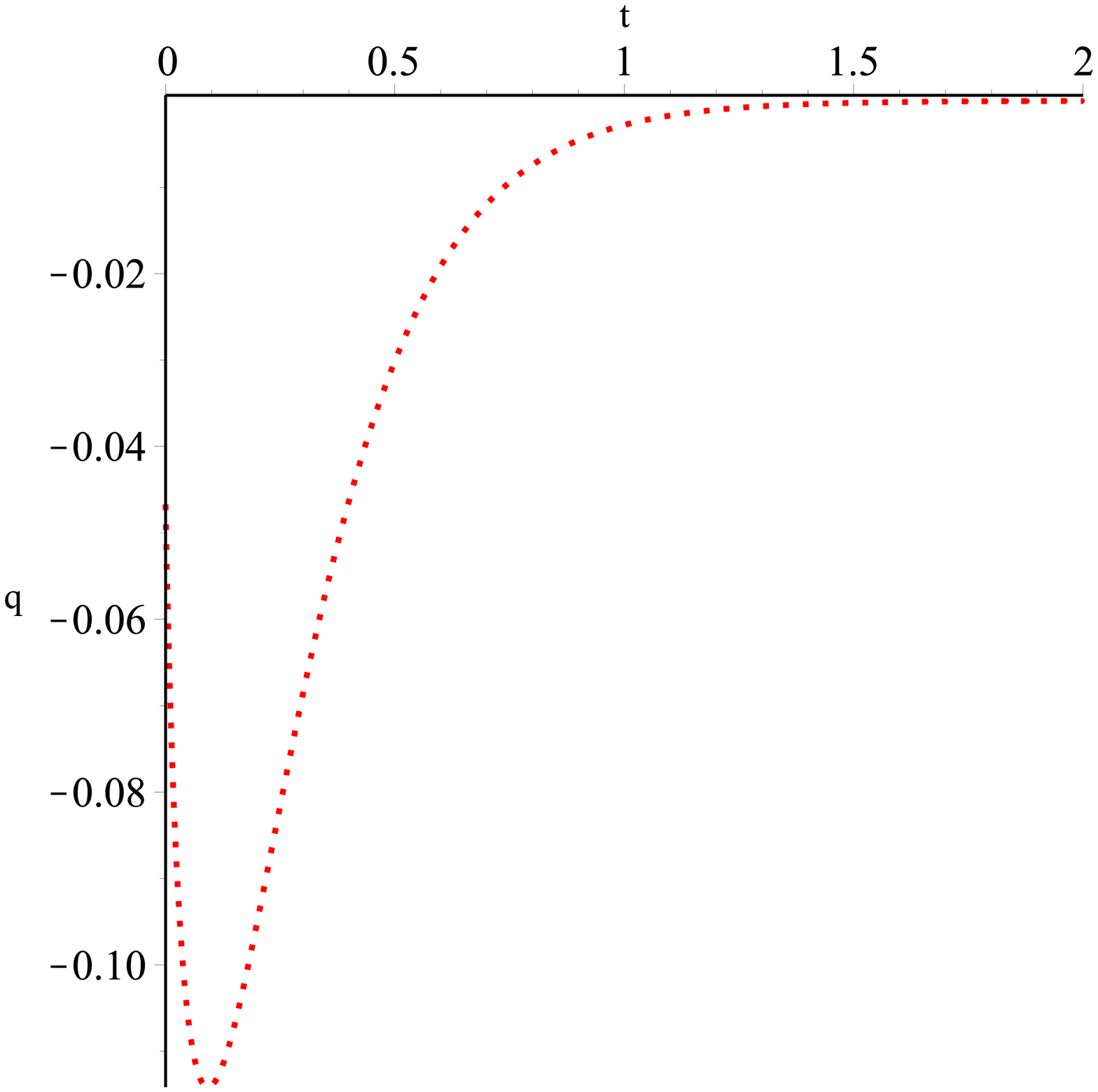} \\  \vskip 1 cm
\caption{Plot of deceleration parameter $q$ with $n_1 = 2/3$, $n_2 =
1/4$, $\lambda_1 = 1$ and $\lambda_2 = 1$} \label{BVIqrqpp}.
\end{figure}

\begin{figure}[ht]
\centering
\includegraphics[height=70mm]{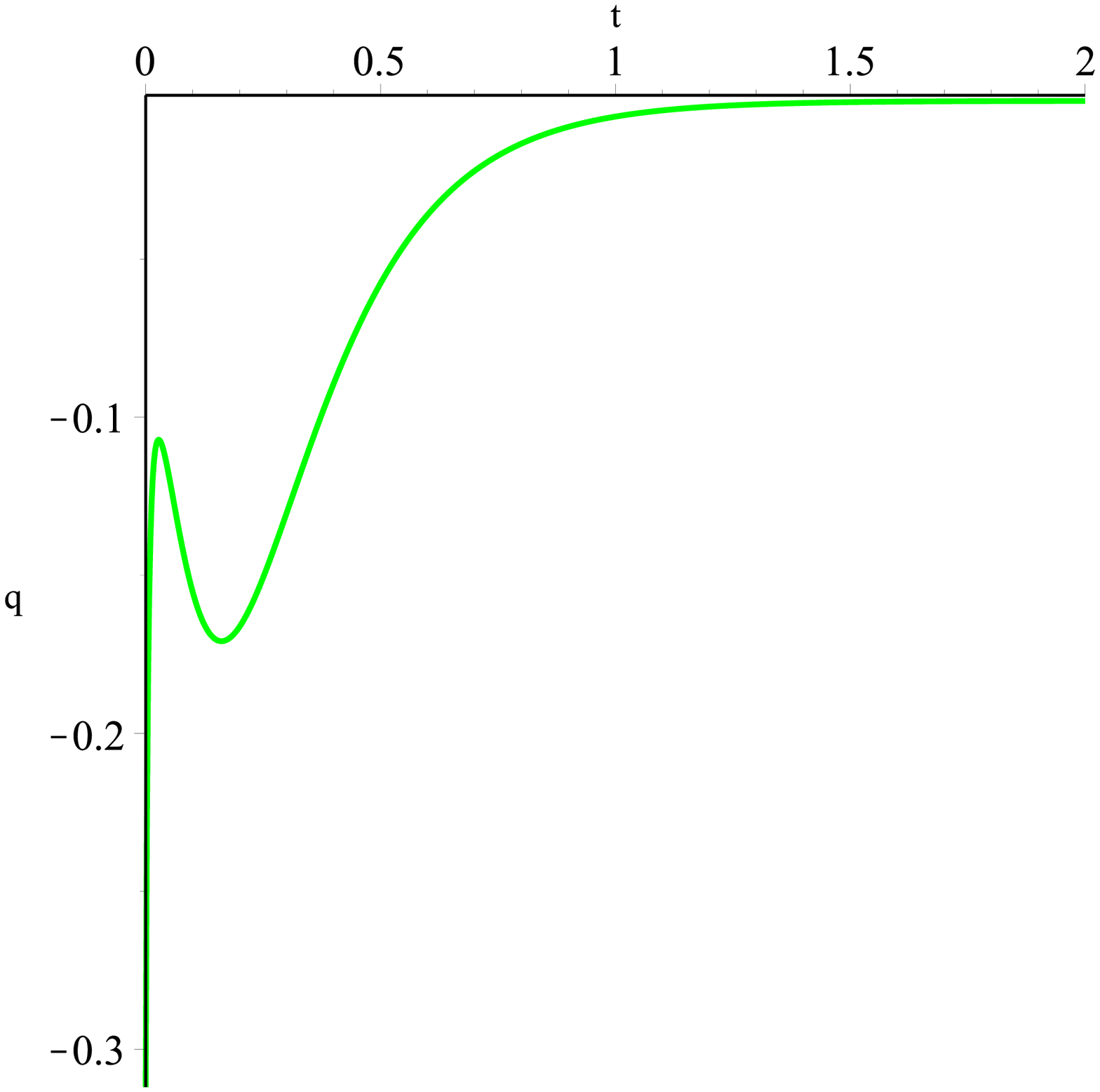} \\  \vskip 1 cm
\caption{Plot of deceleration parameter $q$ with $n_1 = 3/2$, $n_2 =
1/4$, $\lambda_1 =-0.001$ and $\lambda_2 = 1$.} \label{BVIqemqmp}.
\end{figure}

In Fig. \ref{BVIVrqpp} we have illustrated the evolution of volume
scale $V$ for the Universe filled with massive spinor field with
$n_1 = 2/3$, $n_2 = 1/4$, $\lambda_1 = 1$ and $\lambda_2 = 1$. We
draw the picture of evolution of $V$ in the Figs. \ref{BVIVrpmpm},
\ref{BVIVemqmp} and \ref{BVIVempmmm} for $\{ n_1 = 2/3, n_2 = -1,
\lambda_1 = 1, \lambda_2 = -1\}$; $\{n_1 = 3/2, n_2 = 1/4, \lambda_1
=-0.001, \lambda_2 = 1 \}$, and $\{n_1 = 2/3, n_2 = -1, \lambda_1 =
-0.001, \lambda_2 = -1 \}$, respectively. In Figs. \ref{BVIqrqpp}
and \ref{BVIqemqmp} we have illustrated the evolution of
deceleration parameter for positive $\lambda_2$ only.

\section{Conclusion}

Within the scope of Bianchi type-VI spacetime we study the role of
spinor field on the evolution of the Universe. It is found that in
this case the non-diagonal components of the energy-momentum tensor
of spinor field, unlike in the cases Bianchi type I
\cite{sahaAPSS2015} and Bianchi type-$VI_0$ \cite{sahabvi0}, does
not lead to the elimination of spinor field nonlinearity and the
mass term in spinor field Lagrangian. Depending of the sign of
self-coupling constant the model in this case allows either late
time acceleration or oscillatory mode of evolution.

\vskip 0.1
cm

\noindent {\bf Acknowledgments}\\
This work is supported in part by a joint Romanian-LIT, JINR, Dubna
Research Project, theme no. 05-6-1119-2014/2016.


\end{document}